\documentclass[twocolumn]{aastex62}

\graphicspath{{./}}

\received{28 November 2024}
\revised{20 June 2025}
\accepted{1 July 2025}
\submitjournal{AJ}

\shorttitle{Refraction Effects in Kepler}
\shortauthors{Lizotte et al.}

\begin{document}

\title{Exoplanet Atmospheric Refraction Effects in the Kepler Sample}

\author[0009-0002-6280-8681]{Déreck-Alexandre Lizotte}
\affiliation{Department of Physics \& Astronomy, Bishop's University, 2600 Rue College, Sherbrooke, QC J1M 1Z7, Canada}
\affiliation{D{\'e}partement de Physique, Trottier Institute for Research on Exoplanets, Universit{\'e} de Montr{\'e}al, 1375 Th{\'e}r{\`e}se-Lavoie-Roux Av., Montr{\'e}al, QC H2V 0B3, Canada}

\author[0000-0002-5904-1865]{Jason Rowe}
\affiliation{Department of Physics \& Astronomy, Bishop's University, 2600 Rue College, Sherbrooke, QC J1M 1Z7, Canada}

\author[0000-0002-3522-5846]{James Sikora}
\affiliation{Lowell Observatory, 1400 W Mars Hill Road, Flagstaff, AZ, 86001, USA}
\affiliation{Department of Physics \& Astronomy, Bishop's University, 2600 Rue College, Sherbrooke, QC J1M 1Z7, Canada}

\author[0000-0002-1119-7473]{Michael R. B. Matesic}
\affiliation{Department of Physics \& Astronomy, Bishop's University, 2600 Rue College, Sherbrooke, QC J1M 1Z7, Canada}
\affiliation{D{\'e}partement de Physique, Trottier Institute for Research on Exoplanets, Universit{\'e} de Montr{\'e}al, 1375 Th{\'e}r{\`e}se-Lavoie-Roux Av., Montr{\'e}al, QC H2V 0B3, Canada}
\affiliation{D{\'e}partement de Physique, Ciela Institute for Computation \& Astrophysical Data Analysis, Universit{\'e} de Montr{\'e}al, 1375 Th{\'e}r{\`e}se-Lavoie-Roux Av., Montr{\'e}al, QC H2V 0B3, Canada}



\begin{abstract}
We present an analysis on the detection viability of refraction effects in \textit{Kepler}'s exoplanet atmospheres using binning techniques for their light curves in order to compare against simulated refraction effects. We split the \textit{Kepler} exoplanets into sub-populations according to orbital period and planetary radius, then search for out-of-transit changes in the relative flux associated with atmospheric refraction of starlight. The presence of refraction effects---or lack thereof---may be used to measure and set limits on the bulk properties of an atmosphere, including mean molecular weight or the presence of hazes. In this work, we use the presence of refraction effects to test whether exoplanets above the period-radius valley have H/He atmospheres, which high levels of stellar radiation could evaporate away, in turn leaving rocky cores below the valley. We find strong observational evidence of refraction effects for exoplanets above the period-radius valley based on \textit{Kepler} photometry, however those related to optically thin H/He atmospheres are not common in the observed planetary population. This result may be attributed to signal dampening caused by clouds and hazes, consistent with the optically thick and intrinsically hotter atmospheres of \textit{Kepler} exoplanets caused by relatively close host star proximity.
\end{abstract}



\section{Introduction} \label{sec:intro}

Many of the over 5000 confirmed exoplanets so far\footnote{https://exoplanetarchive.ipac.caltech.edu - Accessed on 1/11/23} come with combinations of properties (such as radius, period and composition) that can differ greatly from planetary properties seen in our Solar System. These discoveries have brought about a great deal of changes and new theories regarding our understanding of planetary formation \citep{Ida_Lin_2004, Mordasini_etal_2009, Mordasini_etal_2012, Hansen_Murray_2012, Alibert_etal_2013, Chiang_etal_2013, Chatterjee_Tan_2013, Coleman_Richard_2014, Lee_2014, Raymond_Cossou_2014, Lee_Ciang_2016}. In our own Solar System, there is a divide between the inner and outer planets such that we can divide them into three classes: terrestrial (0.8-1.2 R$_\oplus$), ice giants (4-9 R$_\oplus$) and gas giants (9 R$_\oplus$ and above). Exoplanets appear to show additional classes known as super-Earths (1.2-2 R$_\oplus$) and mini-Neptunes (2-4 R$_\oplus$), which actually make up the bulk of currently discovered exoplanets \citep[e.g.][]{Youdin_2011, Howard_etal_2012, Fressin_etal_2013, Petigura_etal_2013, Morton_Swift_2014, Christiansen_etal_2015, Dressing_Charbonneau_2015}. 

The super-Earth class is characterized by a terrestrial planet (primarily composed of rocky and metallic elements with a solid surface) of near-Earth radius whose atmosphere is small and generally made up of heavier elements. Mini-Neptunes, on the other hand, describe planets with Neptune-like composition (dominated by volatiles, as suggested by their density) but smaller-than-Neptune radii. This means they are generally believed to have solid cores (either rock or ice) with surrounding inflated light (H/He dominated) atmospheres. Further, as reported by recent results from high-resolution spectroscopy, some mini-Neptune atmospheres contain a non-negligible water vapour content (unlike the ice giants in our own Solar System) \citep{Benneke_etal2019, Kite_etal2020, Piette_Nikku2020}.

Some theories of planetary formation predict that both classes would come from a common progenitor population, which forms in-situ and branches out into either class \citep{Lopez_2018}, based on the fact that an observed population valley exists between the two (at around 1.8 R$_\oplus$, for a population that has a period of less than 100 days) \citep{Fulton_2017}, the  {\it period-radius valley}. The progenitor would form within the valley region (or migrate into it), and then either grow large enough to become a mini-Neptune (by accreting H/He, which would inflate the radius without letting the mass reach the critical point for runaway gas accretion \citep{Lopez_2018}), or turn into a super-Earth with its atmosphere evaporated by high radiation from its star. As the \textit{Kepler} host stars are predominantly much older (over 1 Gyr) than the photoevaporation timescale (roughly 100 Myr), we expect a clear division between planets with and without puffier, low mean molecular weight atmospheres. It is based on this theory that we will be exploring if a difference in atmospheric refraction detectability can be observed between the super-Earth and mini-Neptune population in the \textit{Kepler} sample \citep[DR25;][]{Thompson_2018}.

As we observe a planet approaching transit, light from the star will enter the planet's atmosphere, changing medium (from vacuum to an increasingly dense atmosphere) and causing it to be refracted towards the observer. If the atmosphere is optically thin, such that refracted light will eventually reach the observer
the consequent secondary image created by this atmospheric refraction will be crescent-shaped \citep{Misra_2014, Sidis_2010, Dalba_2017, Alp_2018, sheets_2018}. The creation of a secondary image by stellar light will cause the total received flux to slightly increase (proportional to the secondary image's area) and will reach its apparent maximum in the moments before and after the planet's transit. Flux increase due to refraction will reach its true maximum at the mid-transit point (roughly double the out-of-transit maximum increase), however since the signal is competing with the transit itself it is difficult to distinguish. The in-transit refraction effects look like an inverted transit signal and can cause the inferred transit depth to be slightly higher than it really is, leading to degeneracies associated with the transit model.

Previous works \citep{Dalba_2017}(hereafter referred to as D17) have shown that the refraction effect cannot be detected in the photometry of individual \textit{Kepler} close-in ($a <$ 0.2 AU) transiting exoplanets, by using ray-tracing models to estimate the increase in flux due to refraction. These models predict that the mean refraction effect of all \textit{Kepler} exoplanets (regardless of their atmospheric composition) should be on the order of 10 parts-per-million (ppm) (D17). Further, a model based on the methodology of \cite{seagerhui2002}, predicted a similar result of 10 ppm for planets around Sun-like stars (which roughly corresponds to the host stars of\textit{Kepler} targets) \citep{Alp_2018}. By binning photometric light curves together for many systems, the out-of-transit standard deviation was reduced from the 0.1 parts-per-thousand (ppt) to the 1 ppm level. Despite this, the results for binned light curves (for 2394 Kepler Objects of Interest (KOIs) chosen due to the low systematic errors present in their data) from this model once more showed little to no visible refraction effects in the final binned light curve \citep{Alp_2018}. Individual transiting gas giants (chosen where $R_p > 2 R_{\oplus}$, in this case), on the other hand, are shown to have potential refraction effects over the 100 ppm level \citep{sheets_2018} using equations outlined in \cite{Sidis_2010}, which was confirmed by more robust simulated models (D17). 

Our work will build on previous studies by targeting exoplanet populations where atmospheric characteristics are expected to be more favourable to the detection of refraction effects, instead of where orbital parameters predict strong effects. As the exoplanets above the period-radius valley (APRV) are expected to retain lower mean molecular weight atmospheres relative to exoplanets below the valley (BPRV) \citep{Lopez_2012,Owen_2013,Lopez_2018}, their atmospheric scale heights are larger, as is the theoretical cross-section of the visible atmosphere susceptible to refraction. For atmospheres assumed to be made up of an ideal gas in hydrostatic equilibrium, we define the scale height ($H$) as:
\begin{equation}
    \label{scaleheight}
    H = \frac{k_bT_{\rm{atm}}}{\mu g}
\end{equation}
where $k_b$ is the Boltzmann constant, $T_{\rm{atm}}$ is the local atmospheric temperature, $\mu$ is the atmospheric mean molecular mass and $g$ is the local gravitational acceleration (fixed at $GM_{\rm p}/R_{\rm p}^2$). A refraction signal, in this case, would be easier to detect than one coming from an atmosphere, on the same planet, with higher mean molecular weight. These denser atmospheres would present a smaller scale height, and thus present a smaller area where refraction is permissible, causing a refraction signal to be smaller. By binning the light curves of planetary populations that are more likely to have lower mean molecular weight atmospheres and comparing it a similarly binned light curve for the population with low scale heights and a weak refraction effect, we should expect to see a large out-of-transit difference between the two; attributed to the presence of refraction effects.

The works of \citet{Sheets_2014, Sheets_2017} first showed that it is possible to achieve ppm-level photometric uncertainty when binning groups of \textit{Kepler} light curves together. We will similarly bin \textit{Kepler} light curves to achieve a precision of $\sim$1 ppm, with the goal of detecting a difference in out-of-transit refraction effects between the APRV and BPRV exoplanet population. We will present the details of the binning process of \textit{Kepler} photometry, followed by the calculation of simulated refraction effects, as well as injection tests to verify whether refraction effects are dampened and/or removed by the adopted light curve detrending methods. We expand on previous works by splitting the planetary population according to the period-radius valley, which are compared in order to test whether the population of mini-Neptunes shows evidence of refraction and thus, evidence that they are host low mean molecular weight atmospheres. We then compare the binned \textit{Kepler} data with their respective simulated refraction effects to study how they differ. 

\section{Binning Kepler Light Curves} \label{sec:binning}

The photometry used in this work is from the \textit{Kepler} Data Release 25 \citep[DR25;][]{Thompson_2018} catalog and data products. Both confirmed and planetary candidate KOIs from the DR25 and DR25 supplement are included. The observations were processed using the same methodology as presented in DR25 for transit analysis.
We used the pre-search data conditioning (PDC) photometry \citep{keplerpdc}, which is then detrended using a cubic polynomial filter. The detrended light curves are then fit using a $\rm{\chi}^2$ test \citep{Thompson_2018,Rowe_2015}, with a transit model defined assuming circular orbits and fixed quadratic limb darkening coefficients \citep{Mandel_2002}. As the transit parameters from \citet{Thompson_2018} are not best-fit values, but medians of posteriors, we cannot use them to remove the transit model from the data. Instead, we refit the models to find the least square minimum \citep{2016_08_Rowe}. While we have not computed robust statistics for how the best-fit parameters differ from the \citet{Thompson_2018} posteriors, we notice that the change is not relatively significant. Any outliers beyond 5$\sigma$ are clipped. To test if these detrending and clipping processes could affect the presence of out-of-transit signals such as refraction, we performed injection tests to quantify the impact of data reduction (see Section \ref{injectiontests}). It should also be noted that we include (and account for) targets with transit timing variations (TTVs), as well as taking care to discern from transits of other planets in multi-planet systems.

The Kepler sample planets have a range of transit durations such that refraction effects will occur at different times relative to the mid-transit time. Therefore, we first produce phase-folded light curves based on each KOI's orbital phase, which were then rescaled in terms of the transit duration ($T_{14}$). We define the transit duration as the time from first to last contact in the observed light curve, which can be calculated from stellar and planetary parameters \citep{seagerexo}:
\begin{equation}
    \label{tdureq}
    T_{14} = \frac{P}{\pi}\arcsin \left( \frac{R_{\star}}{a} \left[ \frac{\left(1+\frac{R_p}{R_\star}\right)^2 - \left(\frac{a}{R_\star}\cos i\right)^2}{1-\cos ^2i}\right]^{1/2}\right)
\end{equation}
where $P$ is the orbital period, a is the semi-major axis, $R_\star$ is the stellar radius, $R_p$ is the planetary radius and $i$ is the orbital inclination.
This ensures that the out-of-transit portion of each light curve is equally scaled to their respective transit durations, such that they can be overlaid and binned. If the effect is statistically important in many of these light curves, the final binned result should show some part of this effect.

The binning process used a bin size of $0.1\times T14$ (to roughly match \textit{Kepler}'s 30 minute integration time), over a phase range of 6 T14 (from -3 to 3); this is done to ensure enough out-of-transit data is collected (for reference, the simulated effects extend over 1.5 T14). Once that is complete, a weighted average of the flux values is taken:
\begin{equation}
    \label{binningeq}
    W = \frac{\sum^n_id_i w_i}{\sum^n_i w_i}
\end{equation}
where $W$ is the weighted average of each bin (composed of $n$ values), with $i$-th data point $d_i$, whose weight is $w_i = 1/\sigma_i^2$ (where $\sigma_i$ is the uncertainty on $d_i$). The uncertainty for the binned photometric out-of-transit flux ($\sigma_W$) is calculated through propagation of errors as such:
\begin{equation}
    \label{binningeq_sig}
    \sigma_W = \sqrt{\frac{1}{\sum^n_i w_i}}
\end{equation}
As the final binned data is sensitive to systematic noise, a number of KOIs with notably higher photometric error bars were excluded from the binning process: specifically 191/3056 of the APRV population (which contains 25482/398231 individual non-phase folded transits, so 6\%) and 42/583 of the BPRV population (which contains 15175/206477 individual non-phase folded transits, so 7\%) were removed based on the fact that they exhibit residuals with standard deviations $>$1.5 times the expected noise (the mean value of the point-to-point error). This cut-off value was determined by analyzing the distribution of the ratios of standard deviation to expected noise, where a Gaussian-shaped distribution encompassed within the 1.5 limit, with only sparse outliers existing beyond it. The final obtained standard deviation for the resulting binned light curve is on the order of a few parts-per-million---significantly lower than the part-per-thousand level associated with the individual \textit{Kepler} PDC light curves. Therefore, refraction effects with amplitudes larger than several ppm should be detectable in the binned light curve.

\section{Simulated Refraction Effects} \label{sec:klcs}

To model refraction effects we used the \textit{Refraction in Exoplanet Transit Observations} (RETrO)\footnote{https://github.com/pdalba/retro} suite of ray tracing simulations (D17). The flux increase due to refraction depends on the planet's mass ($M_{\rm p}$), semi-major axis ($a$), atmospheric temperature ($T_{\rm atm}$), atmospheric mean molecular weight ($\mu$) and the host star's radius ($R_{\star}$). The planetary mass was estimated from the radius ($R_{\rm p}$) using the mass-radius relationship published by \citet{Chen_2016}. We assume a well-mixed atmosphere in which the nightside and dayside have temperatures given by the equilibrium temperature,
\begin{equation}
    \label{teq}
    T_{\rm{eq}} = T_{\star} (1-\alpha)^{\frac{1}{4}} \sqrt{\frac{R_{\star}}{2a}},
\end{equation}
where $T_{\star}$ is the effective temperature of the host star and $\alpha$ is the planet's Bond albedo, which is assumed to be Earth-like and set to 0.3. While setting T$_{\rm atm} = $T$_{\rm{eq}}$ is an estimate, we use the assumptions based on T$_{\rm{eq}}$ to interpret the excess flux seen before/after transit. 
Of the parameters required by the simulation, only $\mu$ cannot be directly measured or calculated from observables. As such, we will consider it to be the only \textit{free} parameter, as we would otherwise need to make significant assumptions on the planet's atmospheric composition which can change the shape of the refraction effect quite drastically. The other necessary parameters are fit in Section \ref{sec:binning} to create a model refraction effect for each individual KOI.

\subsection{Comparing Photometry to RETrO Simulated Refraction Curves}  \label{subs:mcmc}

\begin{figure}[htb!]
\centering
  \includegraphics[width=\hsize]{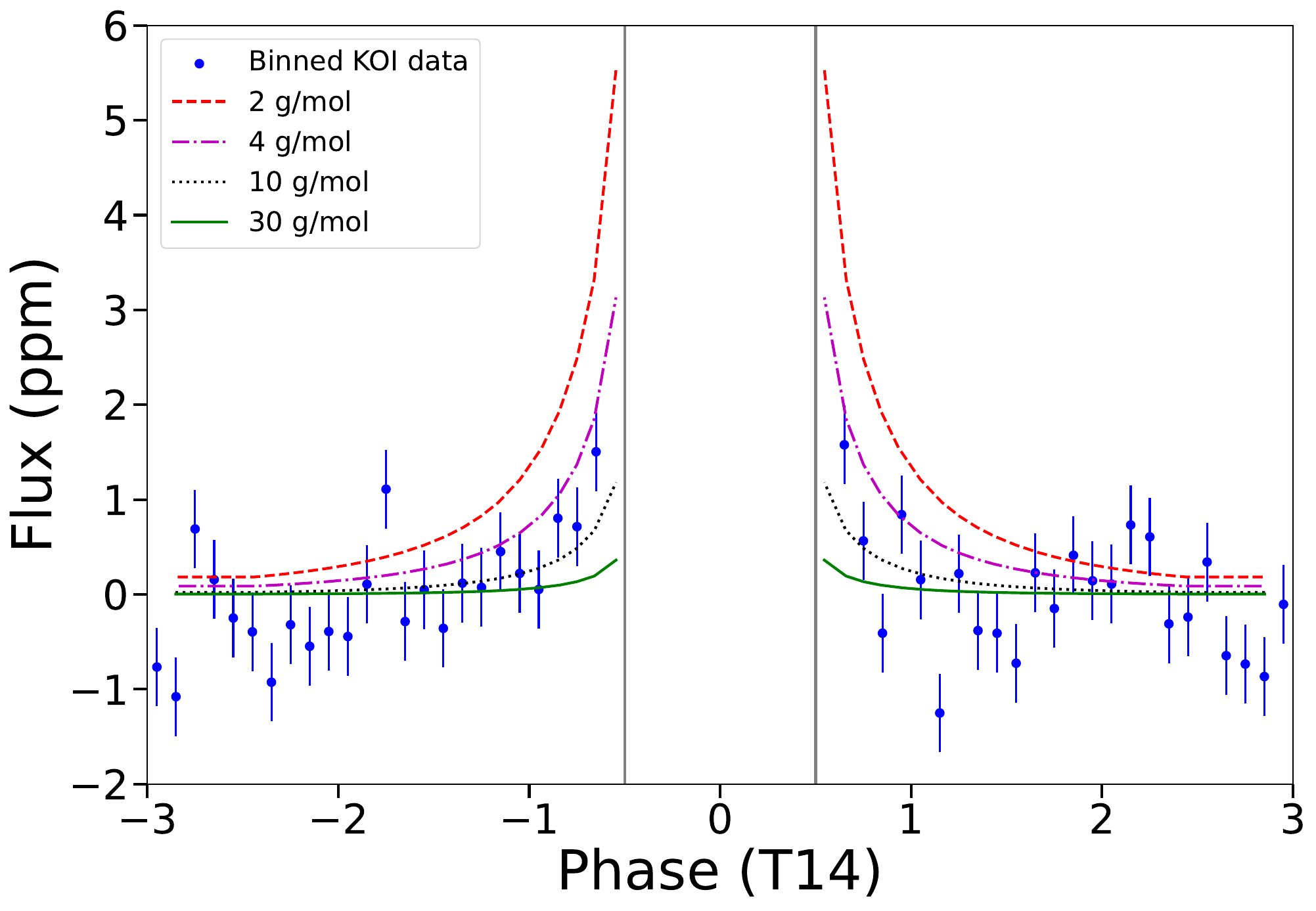} \\
  \includegraphics[width=\hsize]{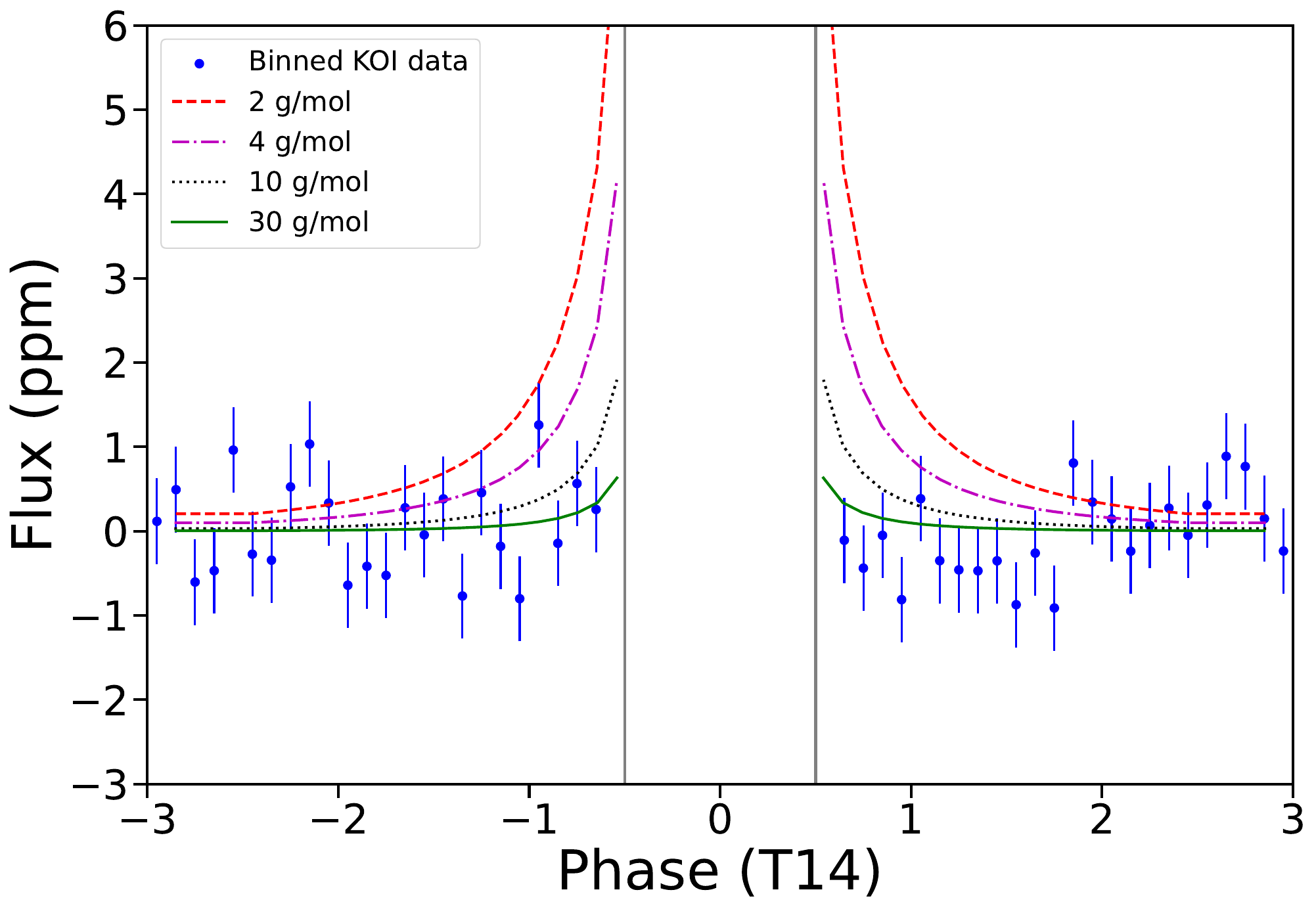}
\caption{Binned \textit{Kepler} photometry (blue dots) for KOI populations above/below the period-radius valley (APRV/BPRV; top/bottom), with refraction simulations overlaid for 2, 4, 10 and 30 g/mol. The black vertical lines indicate ingress and egress, between which all in-transit data has been excised for clarity. Contrary to expectations of the APRV population possessing inflated, low molecular weight atmospheres, the 2 g/mol model atmosphere poorly represents the binned APRV data and is instead better fit between the 4 and 10 g/mol models. The BPRV population behaves as we expected, being relatively flat throughout.}
\label{fig:binkoishoulders}
\end{figure}

In Figure \ref{fig:binkoishoulders}, we compare the binned KOI light curves with the simulated refraction curves generated using different mean molecular weights. The simulated refraction curves were generated independently of photometry and then binned using the same bin widths as for the KOI light curves, such that the the model and data match up. Since we are comparing these simulations with photometric data, it should be noted that \textit{Kepler}'s 30 minute integration time would smooth out refraction effects present in the photometric data. This smoothing will not be present in the simulated data, as it is generated instantaneously (an increase of flux $f$ at a time $t$) at a cadence faster (0.04 T$_{\rm 14}$) than the binned data (0.1 T$_{\rm 14}$). As the photometric out-of-transit flux is not always centered on 0, a  vertical offset was manually applied to the binned transit data, to match its out-of-transit baseline to that of the model. Lastly, the simulated refraction curve was mirrored (around the center of the transit) to compare with the post-transit section of the light curve. 

As the binned refraction curves and binned KOI light curves were overlaid, refraction effects corresponding to low (between 2 and 4 g/mol) mean molecular weight atmospheres were not readily visible (see both parts of Figure\ref{fig:binkoishoulders}), especially when compared to the RETrO simulations. We need only take a cursory glance at Figure \ref{fig:binkoishoulders} to see that the refraction simulation is significantly higher than what the binned light curves produce; this is for the APRV population of exoplanets, where atmospheres with low mean molecular weight (and thus strong refraction) were expected.

We then set out to produce RETrO simulations with different atmospheric compositions (other than just a solar-like H/He composition). The atmospheric mean molecular weight ($\mu$) is important to the maximal flux increase of the simulation, as the atmospheric scale height (Equation \ref{scaleheight}), and by extension the entire refractivity profile, is dependant on it. Further, the changes in $\mu$ will directly be used to interpret any observed refraction signals in the photometric data, as it is the sole dependent parameter within the framework of this project. We consider $\mu = m\times m_{\rm H}$, where $m$ is the molecular weight in g/mol and $m_{\rm H}$ is the atomic mass constant. The atmospheric molecular weights that were chosen for these comparisons were 2 g/mol (corresponding to a H$_2$ dominated atmosphere), 4 g/mol (a He dominated atmosphere), 10 g/mol (a heavier mixture of atmospheric components, used as a half-way point) and lastly 30 g/mol (an Earth-like atmospheric composition). These refraction curves were then compared with the binned transits of both populations (Figure\ref{fig:binkoishoulders}).

With a $\chi^2$ comparison test (with 1 free parameter, $\mu$), we can compare the binned data with each atmosphere type (each already vertically offset to 0), as well as with a null hypothesis (in this case, a horizontal line centered at 0, thus a degree of freedom difference between the null and lensing models). For the above population, the 10 and 30 g/mol model are very similar ($\chi^2$ = 91.5 and 97.2, respectively), with a very marginal favouring of the 10 g/mol model. For the BPRV population, the 30 g/mol regime is favoured over the null hypothesis ($\chi^2$ = 92.3 and 108.1 respectively, so $\Delta\chi^2\sim10$, which implies a significant difference).

\begin{figure}[htb!]
\centering
  \includegraphics[width=\hsize]{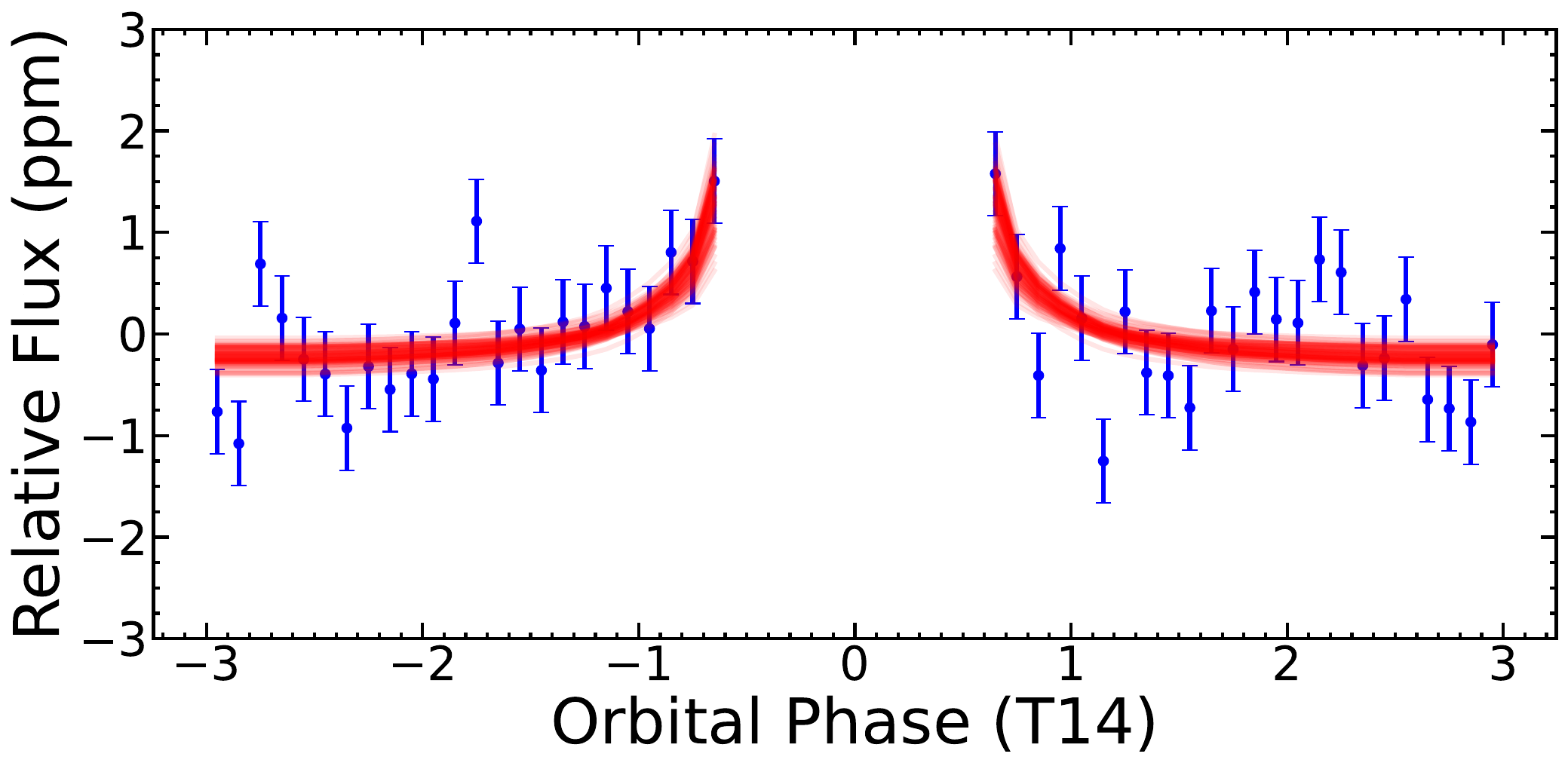} \\
  \includegraphics[width=\hsize]{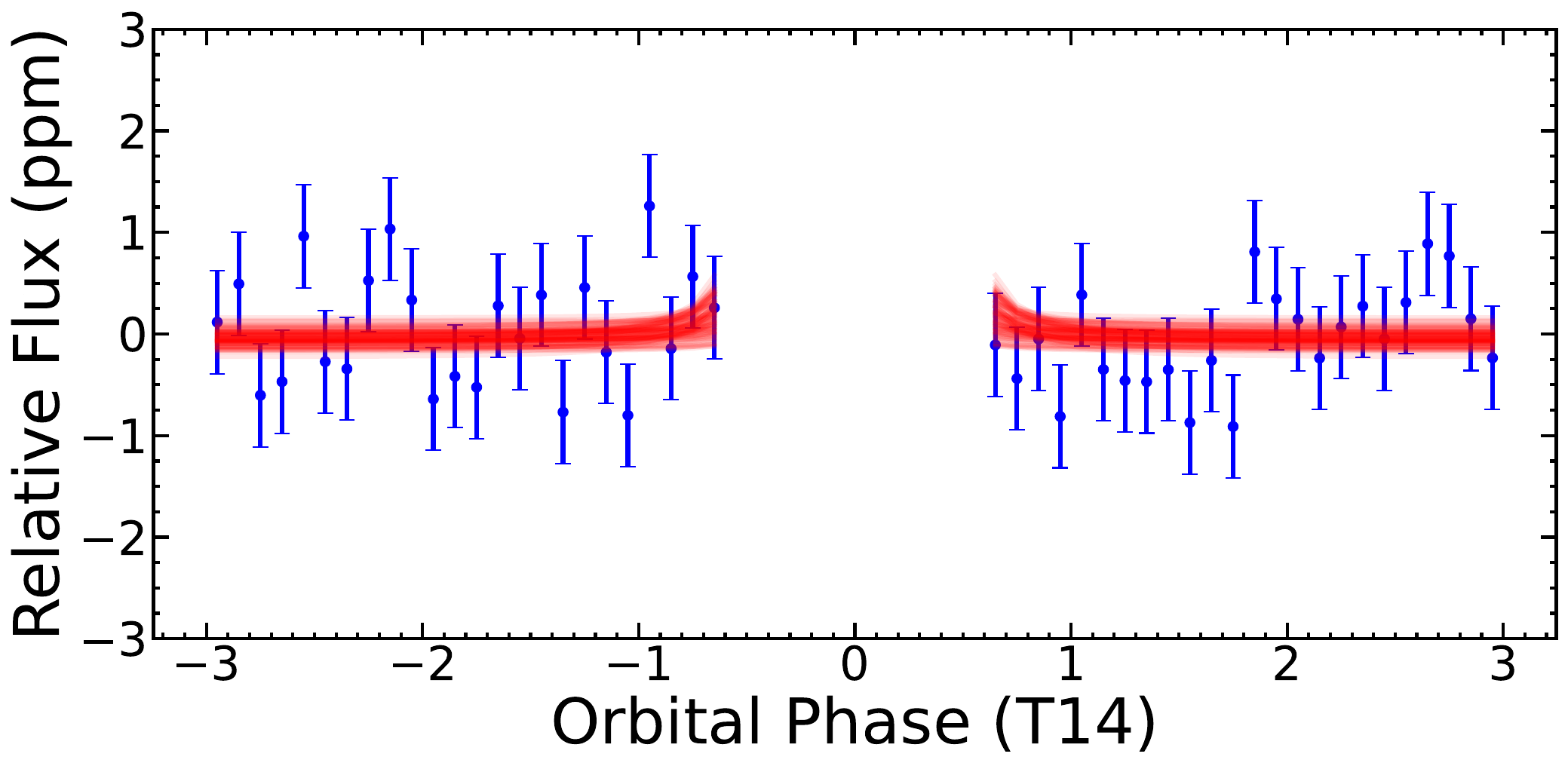}
\caption{\textit{Kepler} photometry (blue dots) for the same above/below (top/bottom) period-radius valley KOI populations as in Figure \ref{fig:binkoishoulders}. Overlaid in solid red lines are 100 models interpolated from RETrO using MCMC posteriors of the zero point ($z$) and mean molecular weight ($\mu$); associated corner plots given in Figure \ref{fig:shouldercorn}.}
\label{fig:shouldergp}
\end{figure}

Lastly, we ran a Markov Chain Monte-Carlo (MCMC) algorithm to fit a RETrO model to the binned KOI light curves. Since running the full simulation as a model in the MCMC sampler would be very expensive computationally, linear interpolation functions were used to fill in the parameter space on $\mu$ between the existing simulated models (2, 4, 10, 30 g/mol). The result (presented in Figure\ref{fig:shouldergp}) showed a slight but clear refraction effect in the APRV population with an amplitude of roughly 2 ppm (corner plots showing the marginalized posteriors are shown in Figure\ref{fig:shouldercorn}). For the BPRV population, the maximum observed amplitude is much closer to 0 ppm, which implies a much heavier atmosphere, likely with a mix of clouds and hazes included, that do not exhibit detectable refraction effects. As the point-to-point scatter of these binned populations remains at slightly below the 1 ppm level, an amplitude of 2 ppm can be hard to discern from systematic noise. To explore if the visible refraction effect may be noise in the data, we divided the APRV population into three seperate groups of 500 randomly chosen KOIs, and applied the same binning process. All three random populations (see Figure\ref{fig:noisetest}) show a slow increase in flux near the transit start/end points, despite the point-to-point noise being greater.

\section{Injection Tests}\label{injectiontests}

\begin{figure}[htb!]
\centering
  \hfill \includegraphics[width=0.965\hsize]{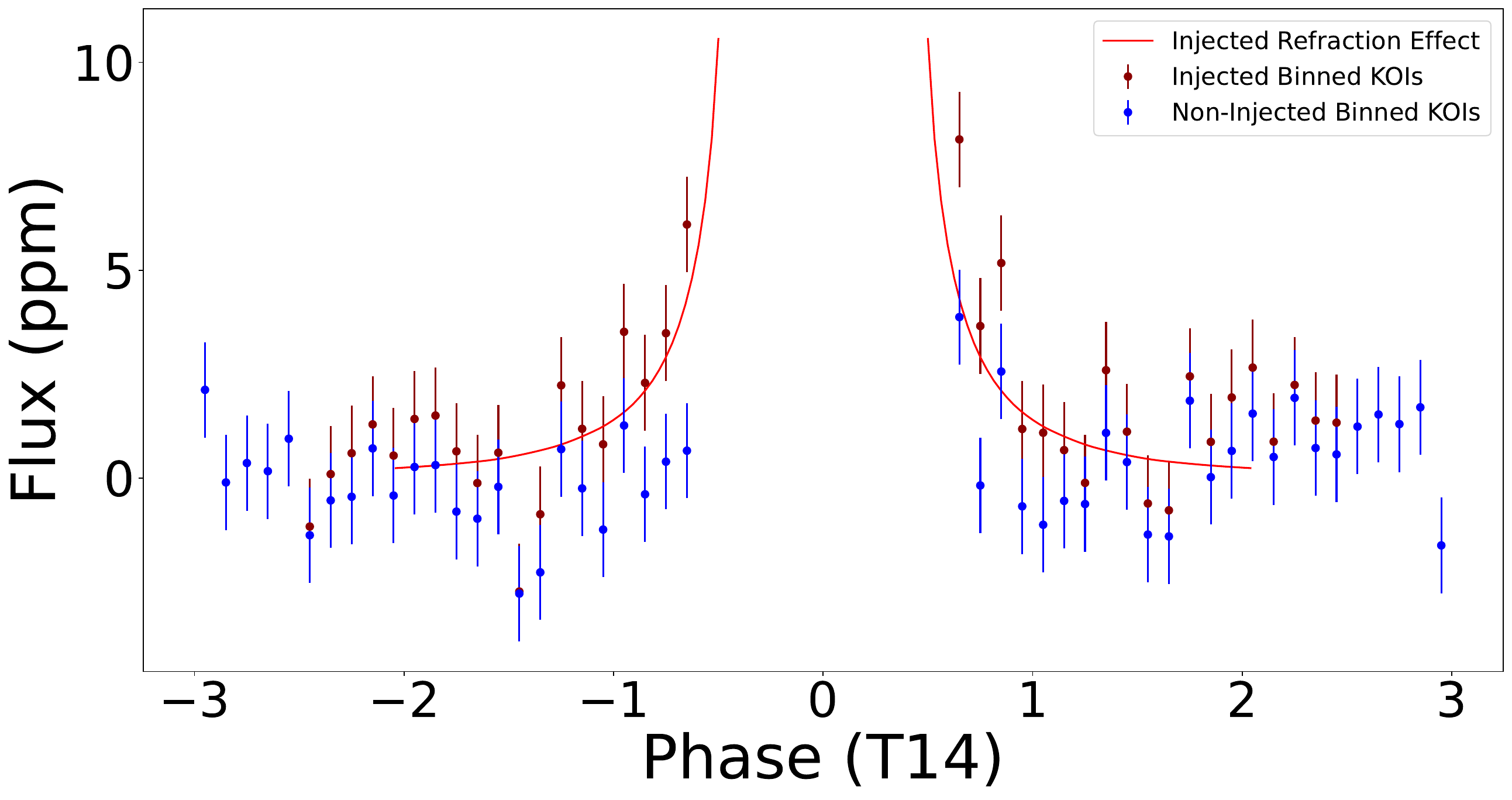} \\
  \includegraphics[width=\hsize]{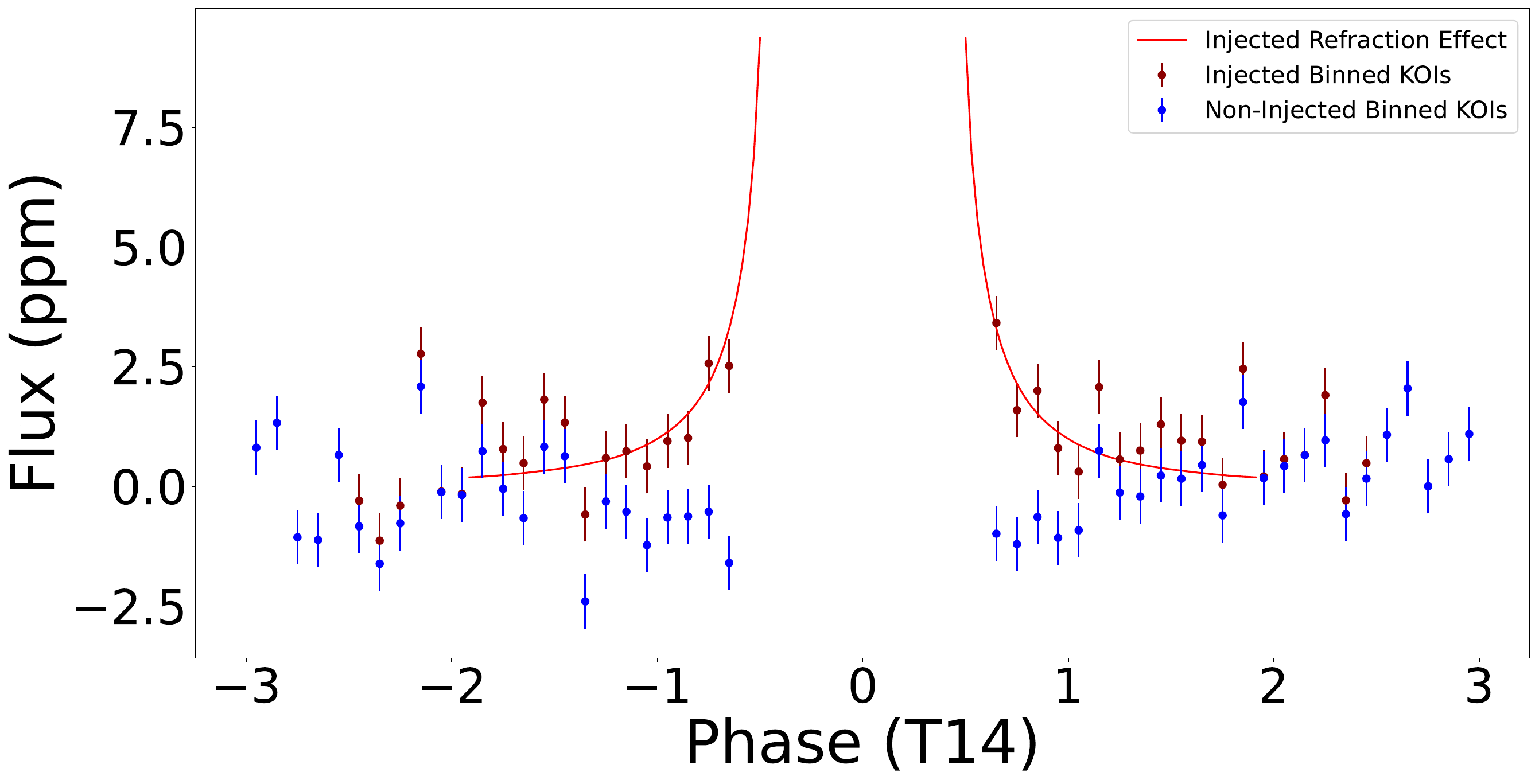}
\caption{Binned \textit{Kepler} photometry of KOI populations above/below the period-radius valley (top/bottom) with/without (red/blue dots) injected refraction signature (solid red line). Prior to binning, a random sample of 300 light curves were injected with a simulated refraction effect---limited to a maximum flux increase of $\sim$10 ppm---in order to assess its significance across the out-of-transit photometry. To protect against the potential for in-transit data to contaminate the out-of-transit refraction signal near ingress/egress by means of \textit{Kepler}'s 30 minute cadence, we exclude data within $\pm 0.1$ T14 from ingress/egress accordingly.}
\label{fig:injectiontestfig}
\end{figure}

The \textit{Kepler} light curves used in the binning procedure have been processed to a certain degree beforehand. This processing includes detrending the raw data, finding transit parameters to create a transit model which is removed from the flux data, and removing statistical outliers ($>5\sigma$) from the residuals. We carried out injection tests in order to determine whether refraction effects that are present in the observed light curves are being impacted by these pre-processing methods (we have not tested against impacts of the \textit{Kepler} pipeline PDC module). This involved injecting synthetic refraction effects generated using RETrO into the raw transit light curves (before the pre-processing). If the final binned light curve still presents clear signs of the injected refracted effect, then we can safely assume that the data processing would not eliminate any real refraction effects in the light curve data. The results of our injection tests can be seen in Figure \ref{fig:injectiontestfig}. The injected increase in flux is visibly reflected in the data, and follows the expected signal (red line).

\section{Discussion}

In comparing the binned KOIs to the simulated effects, we detect a refraction effect consistent with a $\sim 8.8\pm^{2.2}_{0.9}$ g/mol atmosphere in the APRV population, with no detection in the population below. Previous results showed that binned \textit{Kepler} light curves will, in general, only present a flux increase of a few ppm before/after the transit \citep{Alp_2018}, which ends up being indistinguishable from no refraction at all. Our results agree with the first part of this conclusion, with our selection of binned populations allowing us to actually detect a mean refraction effect. However, our results show the APRV population having heavier atmospheres (slightly below 10 g/mol), where we suppose they should have thinner H/He atmospheres (between roughly 2 and 4 g/mol). This presents one of two scenarios: that the atmospheres of some of the considered planetary population are optically thick, or that the APRV exoplanets are composed of heavier mean molecular weight atmospheres (which would be inconsistent with previous observations \citep{Lopez_2018}). In essence, when it comes to using the observed signal of atmospheric refraction effects, an atmosphere with an optically thick low mean molecular weight atmosphere is indistinguishable from one with an optically thin high mean molecular weight.

In the alternate scenario in which the APRV population has a heavier atmospheric composition, we explored the points at which planets can retain certain atmospheric components. Along those lines, there is the idea of a cosmic shoreline \citep{Zahnle_2017}, an approximate relation between a planet's insolation relative to Earth ($I$) and escape velocity ($v_{\rm{esc}}$), which neatly divides the Solar System objects between those with/without atmospheres. This relation would be defined as a power law, as such:

\begin{equation}
    \label{ivescpowerlaw}
    I \propto v_{\rm{esc}}^4
\end{equation}
This relation, along with limits for atmospheres of various chemical compositions (H, H$_2$O, CH$_4$, etc), is neatly plotted in \cite{Zahnle_2017} and also includes some known exoplanets. With this in mind, we set out to plot this power law relation on the KOI population that we considered. Further, we used a classical thermal escape model (Jeans escape) \citep{Zahnle_Catling_2009} for calculating atmospheric escape, for various species of atmospheres. In this calculation, we set planetary temperatures equal to their equilibrium temperature (Equation\ref{teq}) with Bond albedo set to 0.3 (corresponding to Earth's Bond albedo). Using the equilibrium temperature as a temperature of the planet/atmosphere assumes the planet is being heated fully and only by its host star, as if it were a black body. We can consider T$_{eq}$ as being the temperature of the atmosphere at its highest point (R$_p$), where the pressure is essentially zero. Deeper into the atmosphere, the pressure rises as well as the temperature; a particle at higher temperatures could still very well have high enough velocity distribution to escape, despite being at lower altitude. Further, we also must consider setting the Bond albedo to an Earth-like value as a simplification, as planetary reflectivity depends on many unknown factors. Nonetheless, the Bond albedo will have minimal effect (being on the order of unity) when compared to the equilibrium temperature (which is generally on the order of 10$^2$). Our final atmospheric escape lines follow this equation:
\begin{equation}
    \label{ivsvesc}
    I = \frac{v_{\rm{esc}}^2}{108} \frac{\mu}{k_B} \frac{16\pi\sigma}{(1-\alpha)} \frac{a_{\oplus}^2}{L_{\odot}}
\end{equation}
The lines produced by this equation are simplified compared to those in \cite{Zahnle_2017}, which uses a more complex model of atmospheric escape taking into account temperature differences between surface and upper atmospheric layers (for methane-dominated atmospheres, for example) and a more thorough description on the hydrodynamics of atmospheric outflow. The approximate model used in this work considers T$_{\rm{eq}}$ as a lower limit (which realistically would have T$_{\rm{eq}}$ be equal or higher), which translates to the atmospheric escape lines as upper limits.

\begin{figure}[htb!]
    \centering
    \includegraphics[width=\hsize]{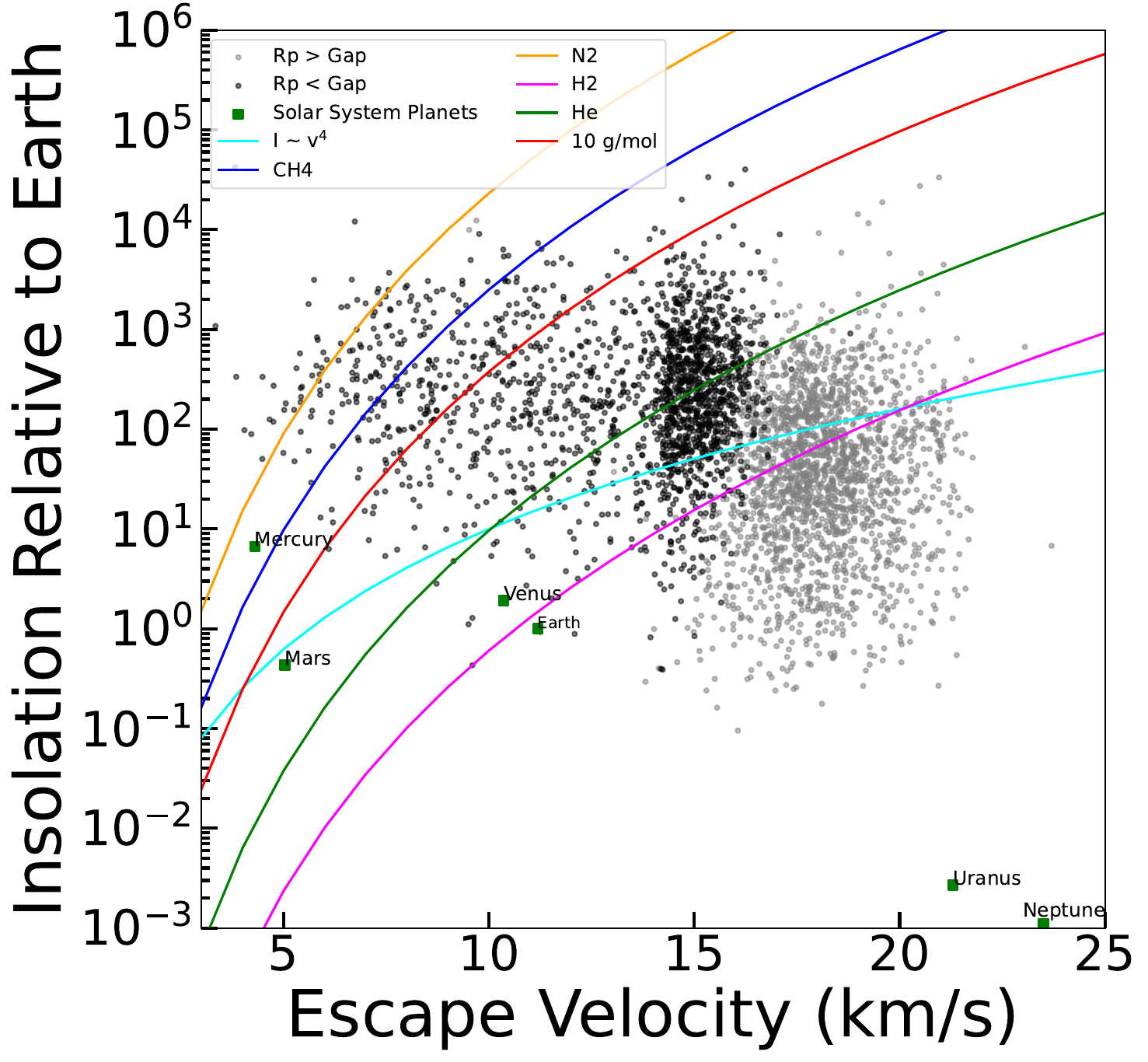}
    \caption{Above/below (grey/black) period-radius valley KOI populations used in the binning process with coloured lines representing various atmospheric escape limits. KOIs below any given lines may viably retain atmospheres of corresponding compositions. Here, the cyan line corresponds to the power law relation between insolation and escape velocity as detailed in \citet{Zahnle_2017}. Solar System planets are also shown for reference. KOI parameters have been updated according to \cite{Lissauer_2024}.}
    \label{fig:fulgapatmoslines}
\end{figure}
From Figure \ref{fig:fulgapatmoslines}, we can see that a sizeable group of the APRV exoplanets are below the H$_2$ atmospheric escape limit. To see whether or not this group has any strong refraction effect (as should be expected if they held H$_2$ dominated atmospheres), we binned this sample and created a RETrO simulation for it where we assume that the atmospheres are H$_2$ dominated.
\begin{figure}[htb!]
    \centering
    \includegraphics[height=5.5cm]{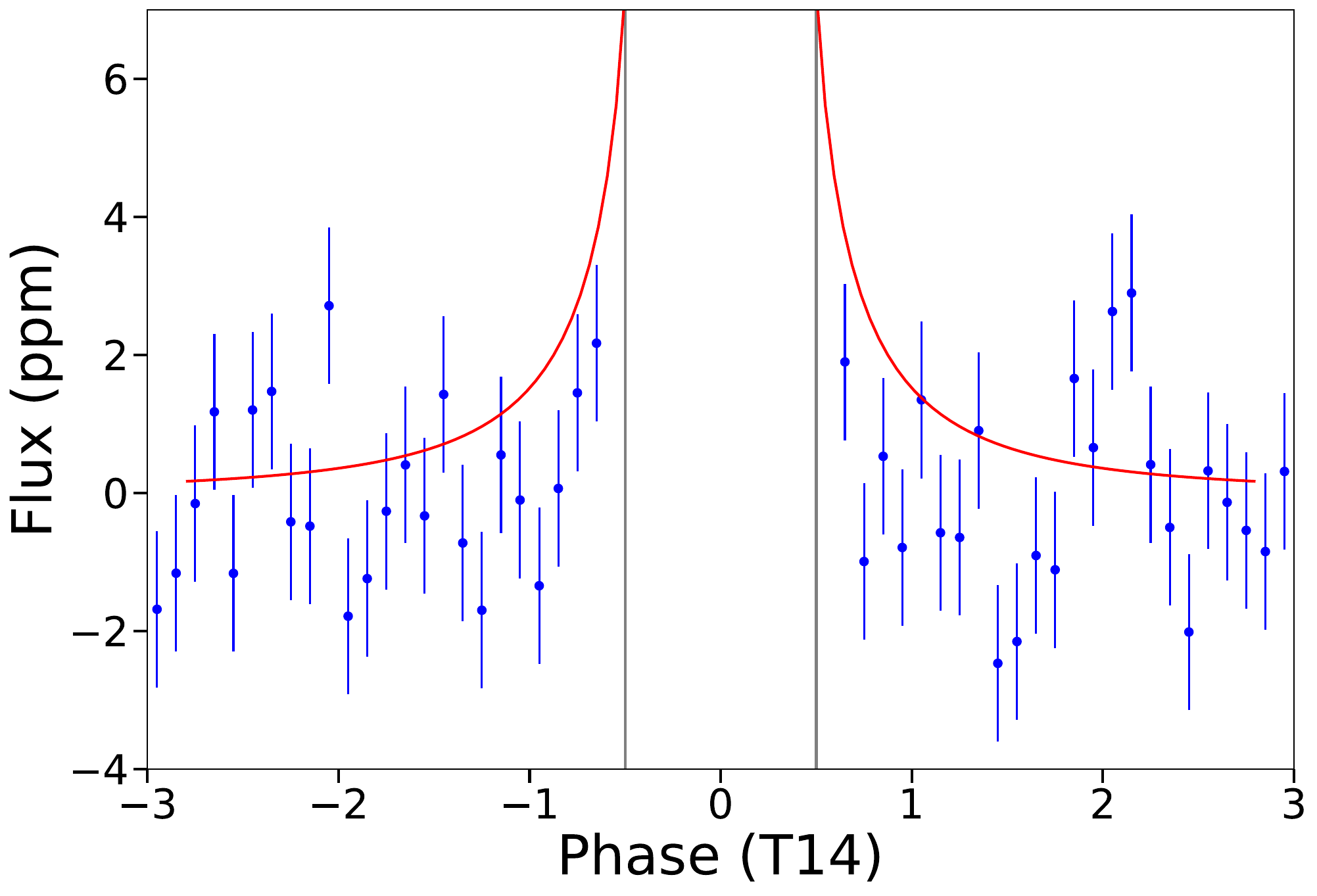}
    \caption{Binned \textit{Kepler} photometry (blue dots) of the KOI population which both exists above the period-radius valley and below the H$_2$ atmospheric evaporation limit shown in Figure \ref{fig:fulgapatmoslines}. Overlaid is a clear H$_2$ atmosphere ($\mu$ = 2 g/mol) refraction model (solid red line). Despite these exoplanets being the only ones in the APRV population capable of holding H/He atmospheres, any possible refraction effects are severely dampened when compared to the simulated effect for an optically thin H/He atmosphere.} 
    \label{fig:binkoishoulderunderh2}
\end{figure}
The result (Figure \ref{fig:binkoishoulderunderh2}) is such that the RETrO simulation predicts a much higher relative flux increase than the binned KOIs report. This is once more indicative of either a lack of H/He dominated atmospheres in this population, or the presence of a dampening factor on the refraction effect, which RETrO does not account for. It is important to note, however, that the approximation that was made for these atmospheric escape lines are such that they are upper limits. In reality, the lines would be much lower, meaning that the actual number of exoplanets below the H$_2$ line (magenta; Figure\ref{fig:fulgapatmoslines}), that is to say the number of exoplanets where a H$_2$ atmosphere will not escape, is much lower than it currently is.

The more likely case is that the \textit{Kepler} sample of exoplanets (both APRV and BPRV) can hold atmospheres which are quite opaque. A cloud layer could completely absorb any light rays passing through the atmosphere, while spotty cloud coverage would partly absorb it. General hazyness in the atmosphere of a planet, or Rayleigh scattering, could also have the same effect or simply reduce the increase in flux from any refraction effects. Previous works have shown that the \textit{Kepler} population of exoplanets are more likely to host hazy atmospheres, which will dampen any refraction effects that are occurring \citep{Misra_2014, Yu_2021, H_rst_2018}. We can consider an optical surface $\tau_{\rm s} = 0.561$ \citep{Lecavelier_des_Etangs_2008, de_Wit_2013}, below which a clear atmosphere is optically thick and no light may pass through. Further, from \cite{betremieux_swain_2017} we can define the effective optical depth ($\tau_{\rm eff}$), which will act as the optical limit from hazes and scattering opacities, as:
\begin{equation}
    \tau_{\rm eff} =  \tau_{\rm s}{\rm e}^{-h/H_{\rm \tau}}
    \label{tau_effective}
\end{equation}
where opacity scale height ($H_{\tau}$) is related to pressure scale height $H$ as $H_{\tau} = H/q$, such that $q$ is a power factor ($q$ equal to 1 for Rayleigh scattering, and 2 for collision-induced absorption opacities). The effective atmospheric thickness $h$ represents the difference between the observed wavelength dependent planetary radius and the optical surface (the value of which is solved in \cite{betremieux_swain_2017}). Finally, it is found that $\tau_{\rm eff} = 0.561{\rm e}^{-E_1(\tau_{\rm s})}$, where $E_1$ is the exponential integral which varies depending on the optical properties of the optical surface and goes to 0 in a clear atmosphere. For a cloud deck, we can simply consider an additional opacity surface at an altitude above the atmosphere's $\tau_{\rm s}$.

\section{Conclusion}

In this study, we binned \textit{Kepler} exoplanet light curves in order to search for weak ($\sim$10 ppm, for a H/He dominated atmosphere) amplitude flux increases caused by atmospheric refraction. We find that in the \textit{Kepler} bandpasses ($\sim400-900$ nm), slight refraction effects are present in some planetary populations but that they are highly dampened when compared to estimated simulations, likely owing to the presence of clouds and hazes in their atmosphere. We detect an average refraction effect in the APRV population that is consistent with a heavier atmospheric mean molecular weight than is expected for optically thin H/He dominated atmospheres of $\mu \in \left[ 2 , 4 \right]$ g/mol. It should also be noted that the orbital periods of both APRV and BPRV populations are typically less than 100 days, which coincide with hotter atmospheric temperatures conducive to hazes. While we cannot directly imply from this conclusion that the period-radius valley splits the \textit{Kepler} planetary population into those with heavier or lighter atmospheres, it does divide it into planetary populations with atmospheres suitable/unsuitable to detectable refraction effects. The question still remains regarding why one side of the divide is more suitable to atmospheric refraction effects, though we can infer this to be likely due to a varying combination of lighter or less opaque atmospheres, possibly also extended over a larger altitude.

As our simulations assume optically thin atmospheres, they do not account for dampening of the refraction effect caused by hazes or clouds. Although this prevents us from reliably investigating atmospheric chemical compositions, it allows us to probe for the presence of atmospheres at a population level across mini-Neptunes and super-Earths. Because Rayleigh scattering---where the intensity $I$ of light at wavelength $\lambda$ scattered by a small particle---follows $I \propto 1/\lambda^4$, we expect the effects of hazes in the refracted beam to drop off significantly at longer wavelengths. Accordingly, it would be advantageous to conduct future atmospheric refraction effect observations in regimes such as the infrared. The James Webb Space Telescope's \citep[JWST;][]{2022jwst} NIRISS instrument is an obvious choice for future infrared follow-up observations, however these would likely be of single exoplanets. Alternatively, the Atmospheric Remote-sensing Infrared Exoplanet Large-survey \citep[ARIEL;][]{arielresume} will also observe in the infrared regime, albeit at a lower resolution. Large population data sets ($\sim$ 1000 exoplanets) of ARIEL would therefore benefit trend-searching efforts similar to this work, especially in terms of atmospheric effects \citep{Helled_2021,Charnay_2021}.

\section*{Acknowledgments} \label{sec:Acknowledgments}

    JFR acknowledges support from the NSERC Discovery program and the Canada Research Chair program. This research made use of Digital Research Alliance of Canada (DRAC; \href{https://alliancecan.ca/en}) computer resources from a RAC allocation to JFR. This research has made use of the NASA Exoplanet Archive, which is operated by the California Institute of Technology under contract with the National Aeronautics and Space Administration through the Exoplanet Exploration Program. The specific observations analyzed can be accessed via \citet{vizierkepler} \dataset[DOI:  10.26093/cds/vizier.22350038]{https://doi.org/10.26093/cds/vizier.22350038}.

\facilities{
    Astrophysics Data System (NASA ADS), NASA
 0.95m Kepler Satellite Mission  \citep{2010_02_Borucki,2010_04_Koch,2016_03_Borucki},
    \href{https://exoplanetarchive.ipac.caltech.edu/index.html}{NASA Exoplanet Archive} \citep{2022_11_NASA_Exoplanet_Archive}
}

\software{
    \texttt{RETrO} \citep{Dalba_2017};
    \texttt{emcee} \citep{Foreman_Mackey_2013};
    \texttt{corner} \citep{2016_06_Foreman_Mackey};
    \texttt{Matplotlib} \citep{2007_05_Hunter};
    \texttt{NumPy} \citep{2020_09_Harris};
    \texttt{pandas} \citep{2010_McKinney,2020_02_The_Pandas_Development_Team};
    \texttt{Python} \citep{1995_01_a_Van_Rossum,1995_01_b_Van_Rossum,1995_01_c_Van_Rossum,1995_01_d_Van_Rossum,1996_05_Dubois,2007_01_Oliphant};
    \texttt{SciPy} \citep{2020_02_Virtanen};
    \texttt{transitfitfive} \citep{2016_08_Rowe};
    \texttt{Astropy \citep{astropy:2013, astropy:2018, astropy:2022}} 
}

\appendix

\section{Additional Figures}

\begin{figure}[htb!]
\centering
  \includegraphics[width=0.47\hsize]{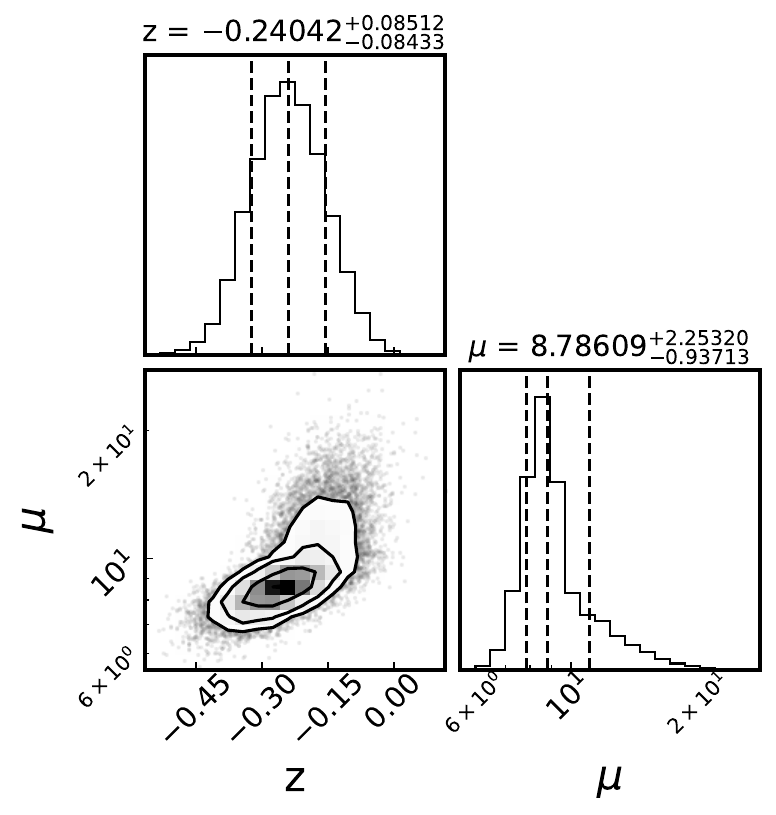}
  \includegraphics[width=0.47\hsize]{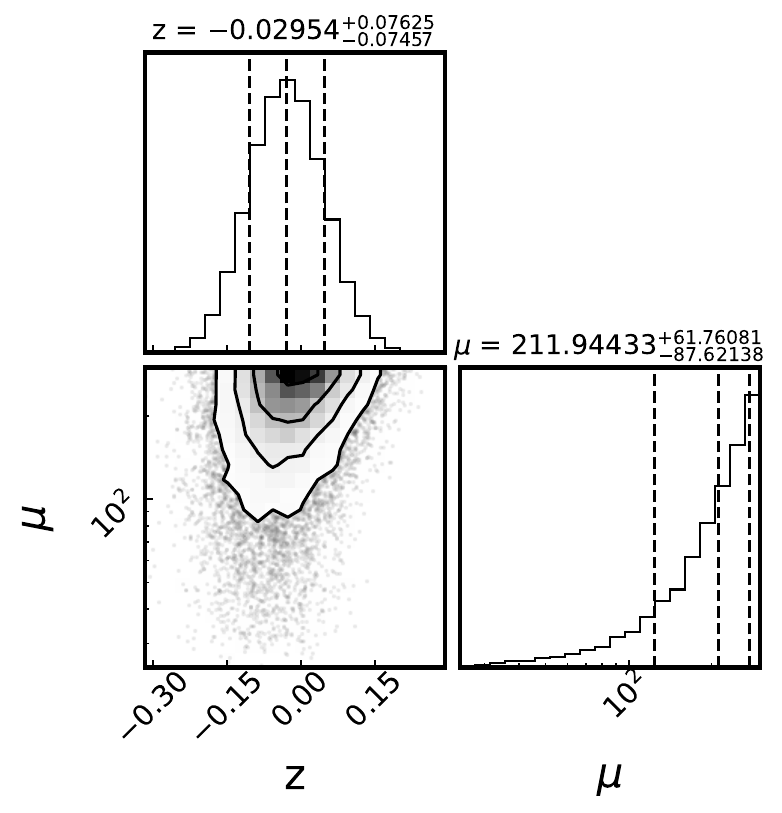}
\caption{Corner plots of MCMC posteriors corresponding to the KOI populations above/below the period-radius valley (APRV/BPRV; left/right), from which samples were drawn for Figure \ref{fig:shouldergp}. The sampled parameters are the zero point ($z$) and mean molecular weight ($\mu$). The dashed lines correspond to a 1-$\sigma$ error. The posterior of $\mu$ associated with the APRV population implies a small, but detectable refraction effect consistent with an optically thin atmosphere lighter than Earth's but heavier than the expected H/He dominated atmosphere. The posterior of $\mu$ associated with the BPRV population implies that when it comes to refraction effects, a much heavier atmosphere is essentially indiscernible from the lack of an atmosphere.}
\label{fig:shouldercorn}
\end{figure}

\begin{figure}[htb!]
\centering
  \includegraphics[width=0.47\hsize]{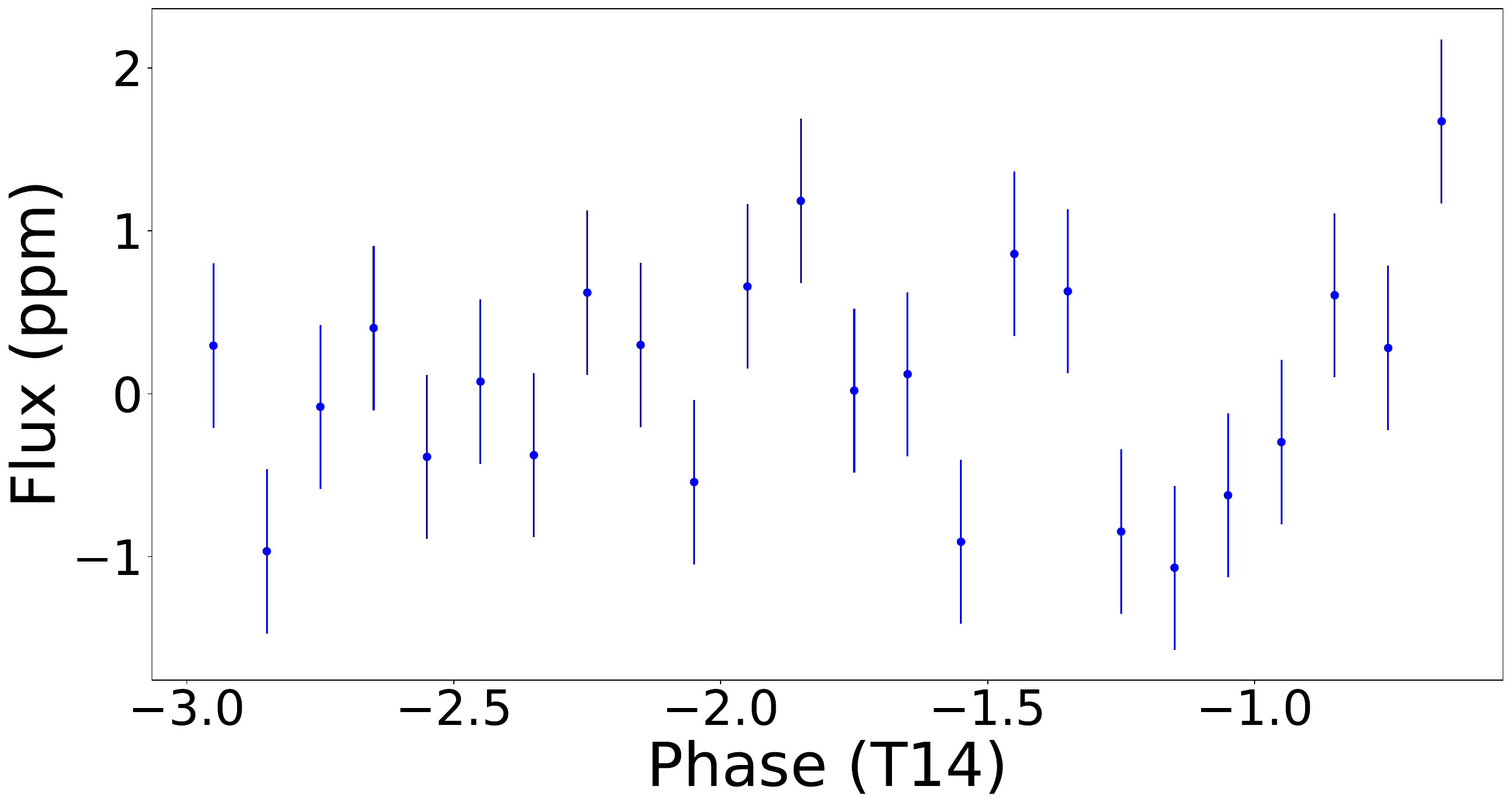}
  \includegraphics[width=0.47\hsize]{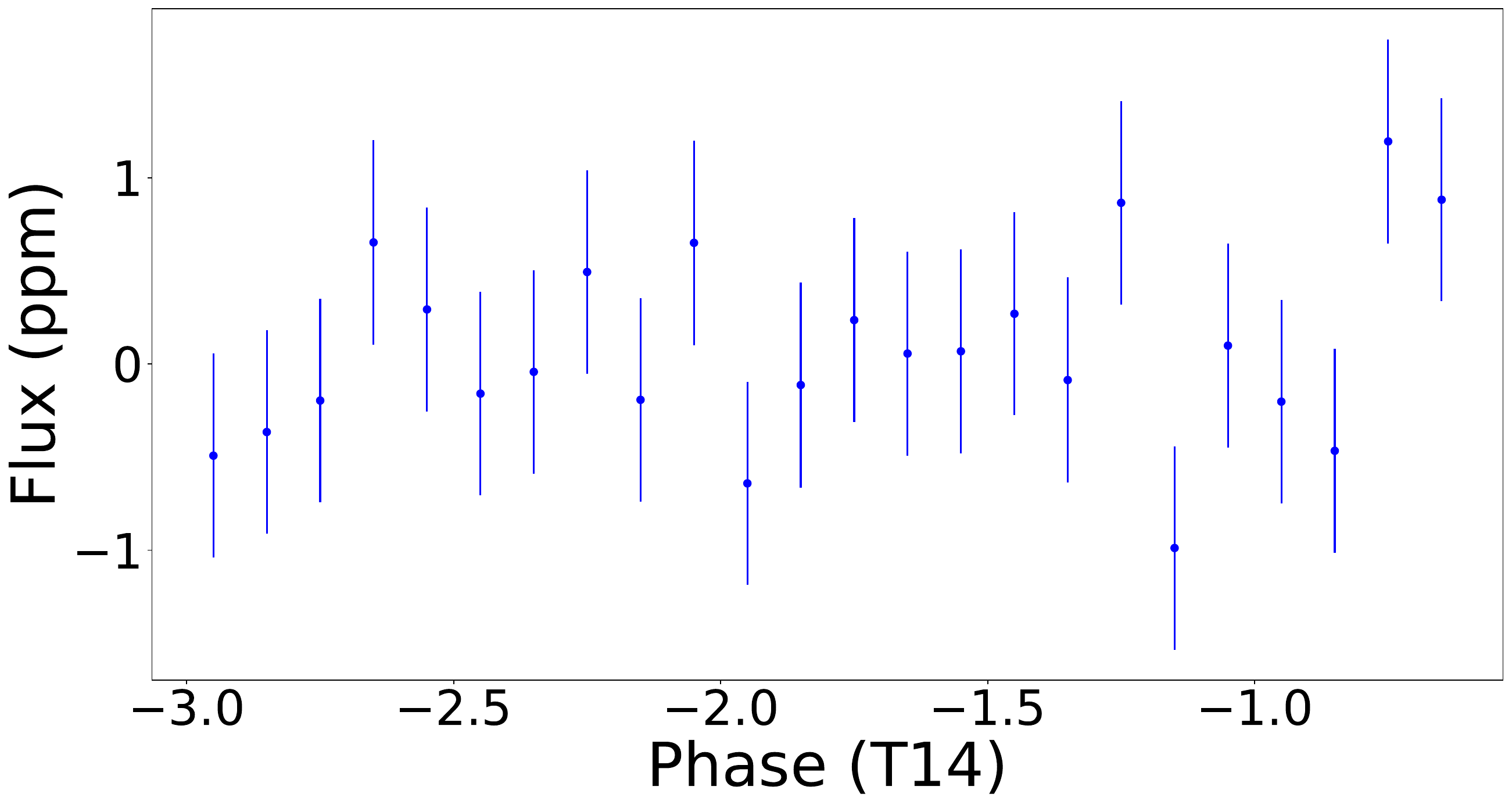} \\
  \includegraphics[width=0.47\hsize]{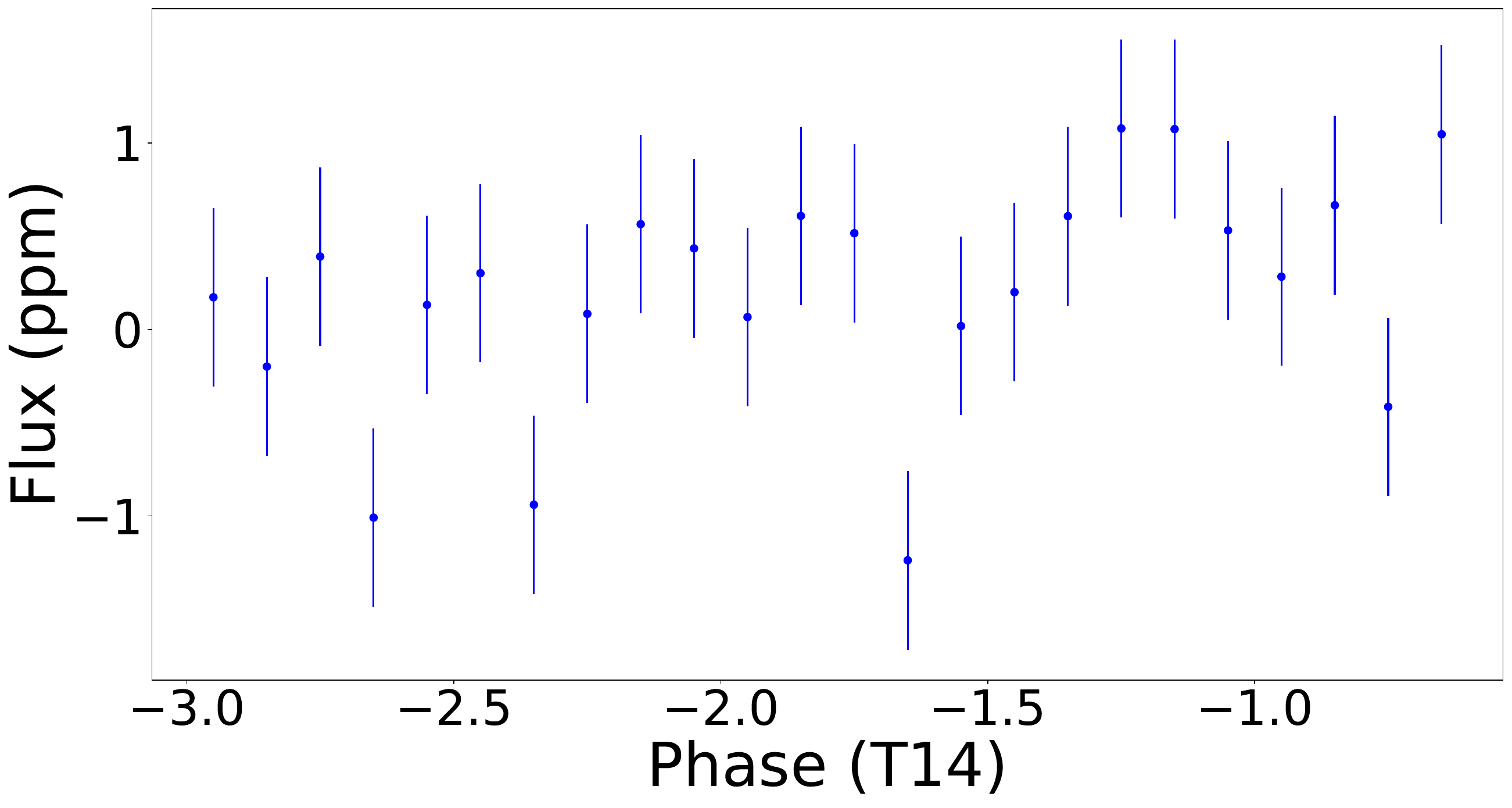}
\caption{Binned \textit{Kepler} photometry (blue dots) corresponding to three 500-KOI realizations (without replacement) from the above period-radius valley (APRV) population. From this, we see significant sensitivity to systematic noise across realizations. As refraction effects are expected to be symmetric, we have binned the pre/post transit data, as to reduce the point-to-point error. Regardless of systematic variability, the refraction effect is distinguishable near ingress/egress in each realization.}
\label{fig:noisetest}
\end{figure}

\clearpage

\section*{ORCID iDs}

    \footnotesize{\noindent Déreck-Alexandre Lizotte \includegraphics[width=9pt]{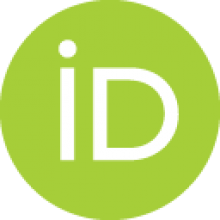} \href{https://orcid.org/0000-0002-1119-7473}{https://orcid.org/0009-0002-6280-8681} \\
    Jason F. Rowe \includegraphics[width=9pt]{Orcid-ID.png} \href{https://orcid.org/0000-0002-5904-1865}{https://orcid.org/0000-0002-5904-1865} \\
    James Sikora \includegraphics[width=9pt]{Orcid-ID.png} \href{https://orcid.org/0000-0002-1119-7473}{https://orcid.org/0000-0002-3522-5846} \\
    Michael R. B. Matesic \includegraphics[width=9pt]{Orcid-ID.png} \href{https://orcid.org/0000-0002-1119-7473}{https://orcid.org/0000-0002-1119-7473}
    }



\bibliographystyle{aasjournal}
\bibliography{biblio.bib}



\end{document}